\documentclass[runningheads]{llncs}
\usepackage[T1]{fontenc}
\usepackage[pdftex]{graphicx}
\usepackage{floatflt}
\usepackage{float}
\usepackage{amsmath}
\usepackage{orcidlink}
\usepackage[
backend=biber,
style=numeric, maxbibnames=20,
sorting=none, doi=false
]{biblatex}
\usepackage{mathtools}

\usepackage{array}
\newcolumntype{M}[1]{>{\centering\arraybackslash}m{#1}}
\newcolumntype{N}{@{}m{0pt}@{}}

\begin{document}

\title{Second 6G life Workshop on\\ Post Shannon Theory}

\author{Yaning Zhao\inst{1,2}\orcidlink{0009-0003-3601-5503} \and
Christian Deppe\inst{1,2}\orcidlink{0000-0002-2265-4887}}
\authorrunning{Y- Zhao, C. Deppe}
%
\institute{{
Technical University of Braunschweig, Institute for Communications Technology, Braunschweig, Germany \and 6G-life, 6G research hub, Germany} \\
\email{yaning.zhao@tu-bs.de, christian.deppe@tu-bs.de}}

\maketitle

\centerline{In Memory of Ning Cai}

\begin{abstract}
The one-day workshop, held prior to the "ZIF Workshop on Information Theory and Related Fields", provided an excellent opportunity for in-depth discussions on several topics within the field of post-Shannon theory. The agenda covered deterministic and randomized identification, focusing on various methods and algorithms for identifying data or signals deterministically and through randomized processes. It explored the theoretical foundations and practical applications of these techniques.
The session on resources for increasing identification capacity examined the different resources and strategies that can be utilized to boost the capacity for identifying information. This included discussions on both hardware and software solutions, as well as innovative approaches to resource allocation and optimization.
Participants delved into common randomness generation, essential for various cryptographic protocols and communication systems. The session highlighted recent advancements and practical implementations of common randomness in secure communications.
The workshop concluded with a detailed look at the development and practical deployment of identification codes. Experts shared insights on code construction techniques, implementation challenges, and real-world applications in various communication systems.
We extend our thanks to the esteemed speakers for their valuable contributions: Caspar von Lengerke, Wafa Labidi, Ilya Vorobyev, Johannes Rosenberger, Jonathan Huffmann, and Pau Colomer. Their presentations and insights significantly enriched the workshop.
Additionally, we are grateful to all the participants whose active engagement, constructive comments, and stimulating discussions made the event a success. Your involvement was crucial in fostering a collaborative and intellectually vibrant environment.
\end{abstract}
\sloppy
\section{Introduction}

As numerous experts in the field of post-Shannon theory had registered for the memorial workshop honoring Ning Cai at the Center for Interdisciplinary Research (ZiF) in Bielefeld, a unique opportunity emerged to further capitalize on the presence of these specialists. It was decided to organize a second workshop focused on post-Shannon theory as part of the 6G-life project. This additional workshop was scheduled to take place at Bielefeld University the day before the memorial event.

The organization of this pre-event workshop was made possible thanks to Jens Stoye, who facilitated the logistical arrangements by providing a suitable room at the university. His support enabled Christian Deppe to seamlessly arrange the workshop, ensuring a conducive environment for productive discussions and exchanges.

The workshop was met with considerable enthusiasm and was well attended by both leading experts in post-Shannon theory and participants from related fields. Interestingly, it also attracted several attendees who were not directly involved in post-Shannon theory research but were keen to learn and contribute to the discussions. This diverse participation fostered a rich exchange of ideas, making the workshop a resounding success and setting a collaborative tone for the subsequent memorial workshop.
\
\section{Workshop Talks}

The concept for the workshop was to utilize a tutorial that had been initially developed for the VCC 2023 (IEEE conference). This tutorial would serve as a foundation to cover key topics, followed by in-depth discussions among the participants. Christian Deppe began by welcoming everyone and providing an overview of the workshop's structure and objectives. He then introduced Caspar von Lengerke, who was responsible for leading the first session.
\newpage
\subsection{Caspar von Lengerke}
\begin{floatingfigure}[r]{6cm}
\mbox{\includegraphics[width=5.5cm]{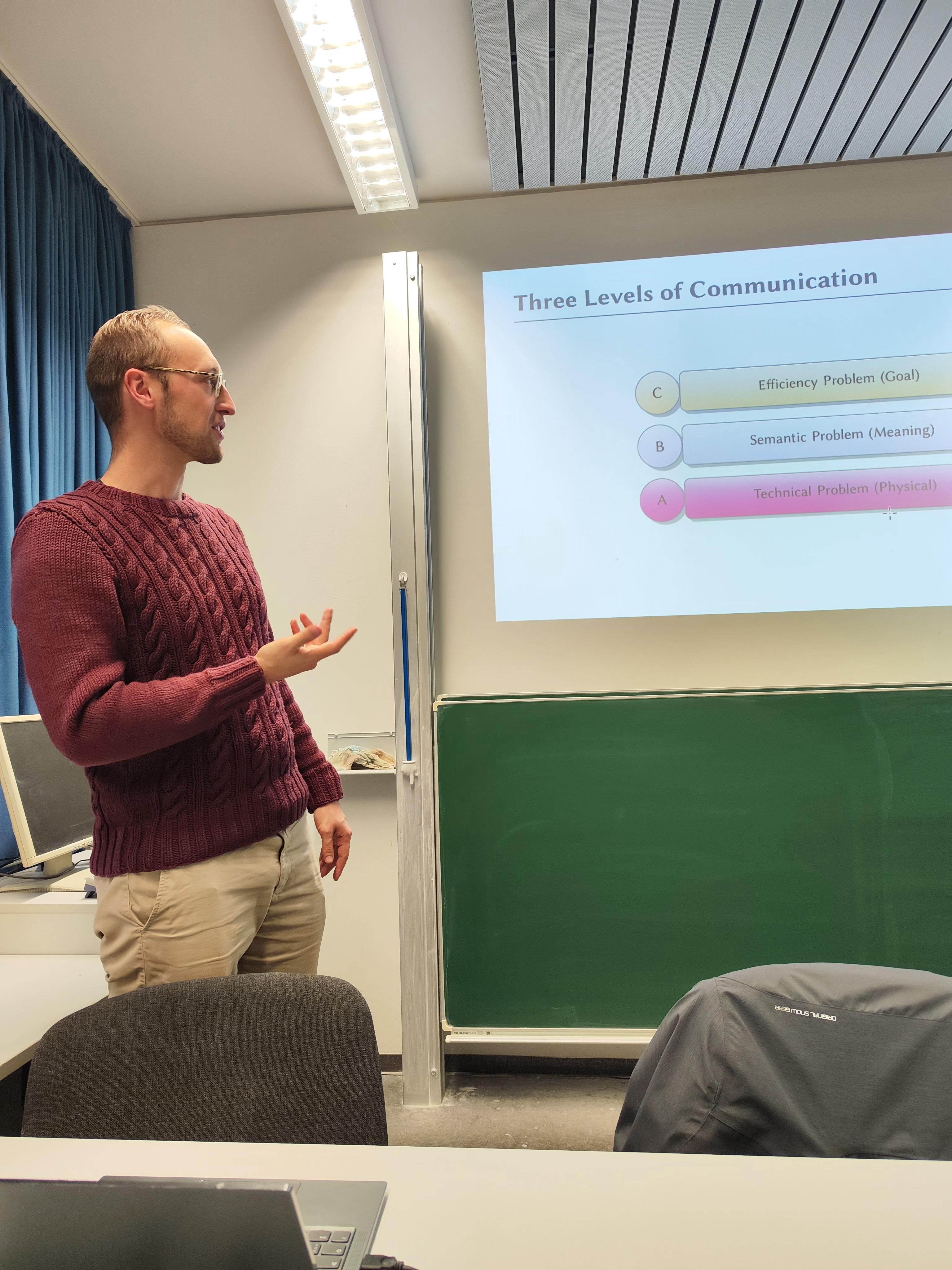}}
\caption{Caspar von Lengerke}
\end{floatingfigure}
In this initial part of the workshop, Caspar von Lengerke delved into the topic of message identification. He started by defining randomized identification codes, laying the groundwork with the fundamental theoretical results established by Ahlswede and Dueck \cite{ahlswede1989identification}. Caspar then moved on to discuss more recent advancements in the theory, providing the audience with a comprehensive overview of the field's evolution.
Caspar’s own research focuses on the practical implementation of message identification systems \cite{von2023beyond,caspar3}. He shared insights from his work, highlighting innovative ideas on how message identification codes can be effectively used in real-world applications \cite{derebeyouglu2020performance,ferrara2022implementation,caspar1,caspar2}. His presentation sparked a lively and engaging discussion among the participants, who contributed their perspectives and questions.
To conclude his session, Caspar directed the participants' attention to his poster, which showcased his latest findings and was on display at the memorial workshop. This poster provided additional details and visual representations of his research, inviting further exploration and dialogue.

\newpage
\subsection{Wafa Labidi}
\begin{floatingfigure}[r]{6cm}
\mbox{\includegraphics[width=5.5cm]{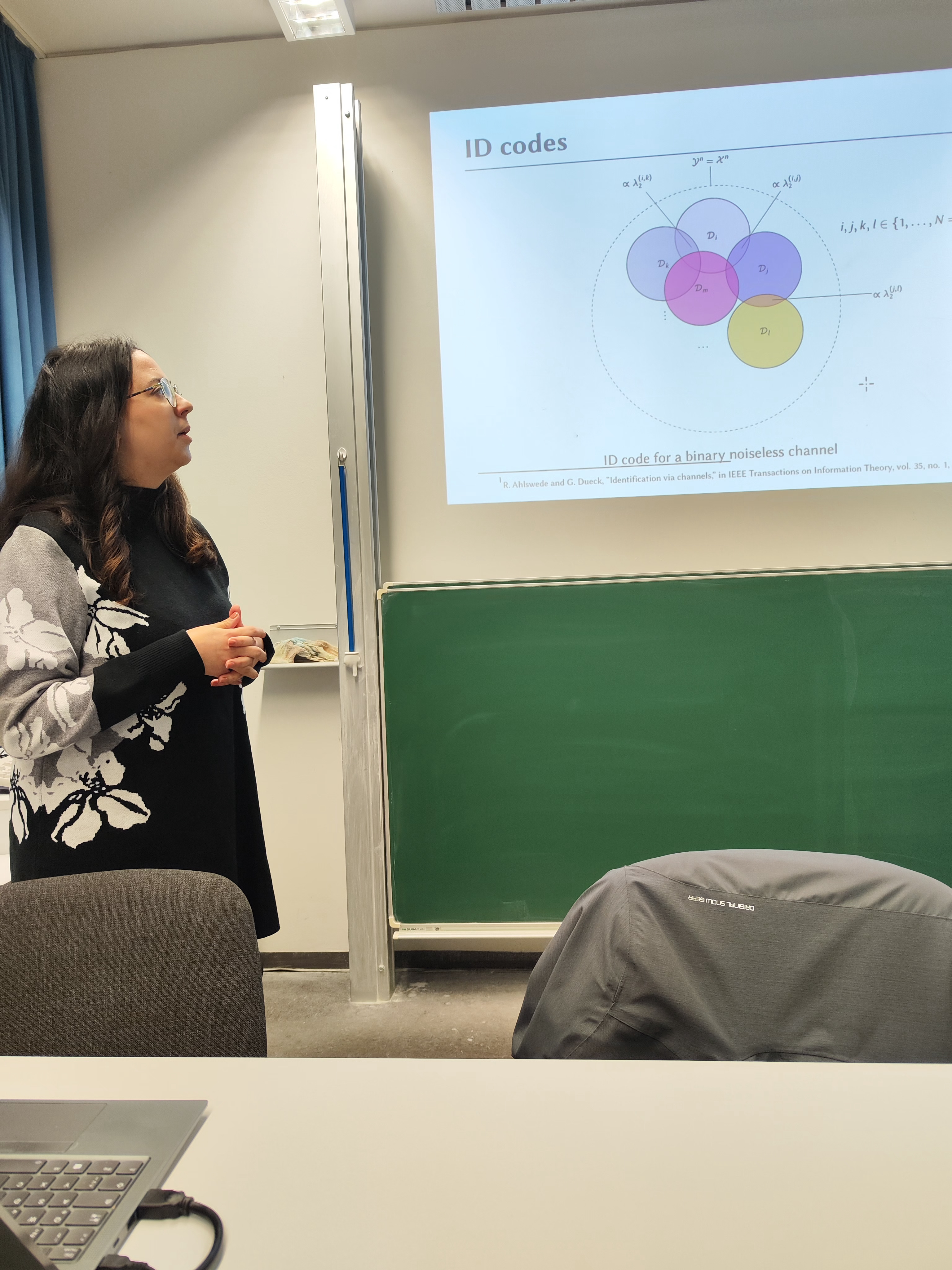}}
\caption{Wafa Labidi}
\end{floatingfigure}
Following Caspar's presentation, Wafa Labidi took the stage to continue the workshop. Caspar had already highlighted the significance of common randomness generation in the context of message identification, noting its role in enhancing identification capacity. Building on this foundation, Wafa began her session by elaborating on the model of common randomness generation \cite{ezzine2020common,labidi2022common,labidi2023common}. She then transitioned into a discussion on correlated assisted message identification, explaining how these concepts interlink to optimize identification processes.
One of Wafa’s areas of expertise lies in working with continuous alphabets \cite{labidi2023common}. She presented her findings related to common randomness generation and its application in message identification within this framework \cite{ezzine2020common}. Her discussion was rich with detailed theoretical insights and practical implications, which sparked numerous questions and comments from the audience, leading to a vibrant exchange of ideas.
Wafa also introduced feedback as another valuable resource for message identification \cite{wiese2022identification}. She discussed how incorporating feedback mechanisms can further enhance the identification process, offering examples and theoretical backing for her points \cite{labidi2023joint,zhao2024achievable}.
Expanding the scope of the discussion, Wafa explored the intersection of molecular communication and message identification. She explained how concepts from message identification are applicable to molecular communication systems. Delving deeper, she described the relationship between the Poisson channel—a fundamental model in molecular communication—and its relevance to message identification \cite{salariseddigh2023deterministic}.
Wafa then shared her own research results on randomized message identification via the Poisson channel \cite{labidi2023information}. Her presentation included detailed explanations of her methodologies, findings, and the implications of her work for both theoretical research and practical applications. The session concluded with a robust discussion, where participants engaged with Wafa’s insights, asking questions, and offering their own perspectives.

\newpage
\subsection{Ilya Vorobyev}
\begin{floatingfigure}[r]{6cm}
\mbox{\includegraphics[width=5.5cm]{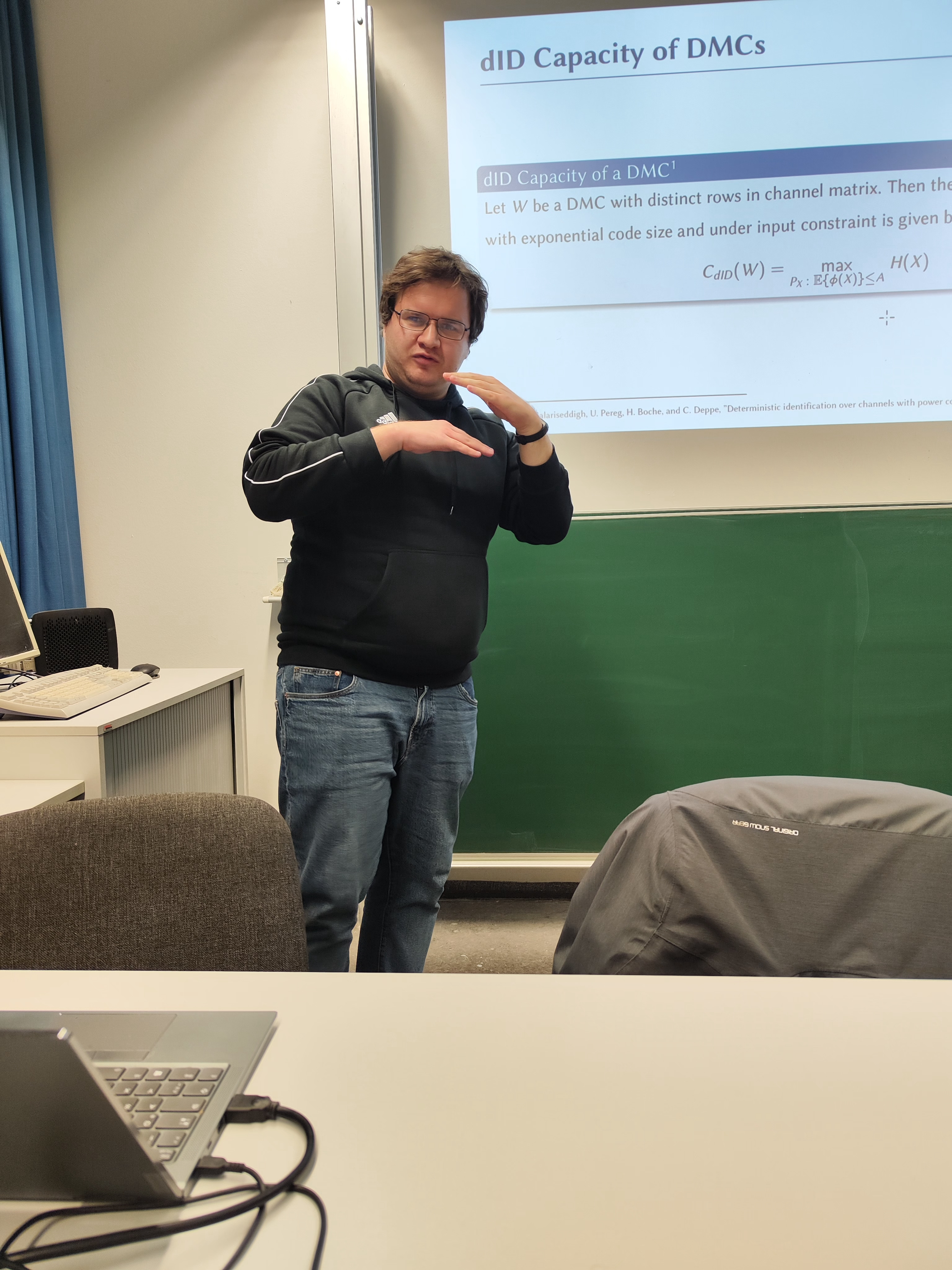}}
\caption{Ilya Vorobyev}
\end{floatingfigure}

After Wafa's presentation, Ilya Vorobyev took the stage. In his detailed presentation, he delved into the intricacies of deterministic identification, a topic he has extensively researched \cite{vorobyev2024deterministiccode}. One key aspect of deterministic identification, as Ilya explained, is the assumption that the sender cannot utilize local randomization for encoding the messages. This assumption sets deterministic identification apart from other methods. Ilya built upon Wafa's point that deterministic identification outperforms traditional transmission codes in certain scenarios. However, this performance boost comes with a trade-off: the typical double-exponential growth rate seen in randomized coding strategies is no longer achievable.
To illustrate his points, Ilya introduced his own construction of deterministic identification codes \cite{vorobyev2024deterministic}. He compared these with Caspar's well-known randomized code construction \cite{von2023beyond}, highlighting both the strengths and weaknesses of his deterministic approach.
A significant part of Ilya's presentation was dedicated to his recent findings on deterministic identification in fading channels \cite{vorobyev2024deterministic}. He had derived new bounds for this scenario just before the workshop, marking a fresh contribution to the field. These findings prompted a lively and engaging discussion among the workshop participants, who were keen to explore the implications and potential applications of Ilya's work.

\newpage
\subsection{Jonathan Huffmann}
\begin{floatingfigure}[r]{6cm}
\mbox{\includegraphics[width=5.5cm]{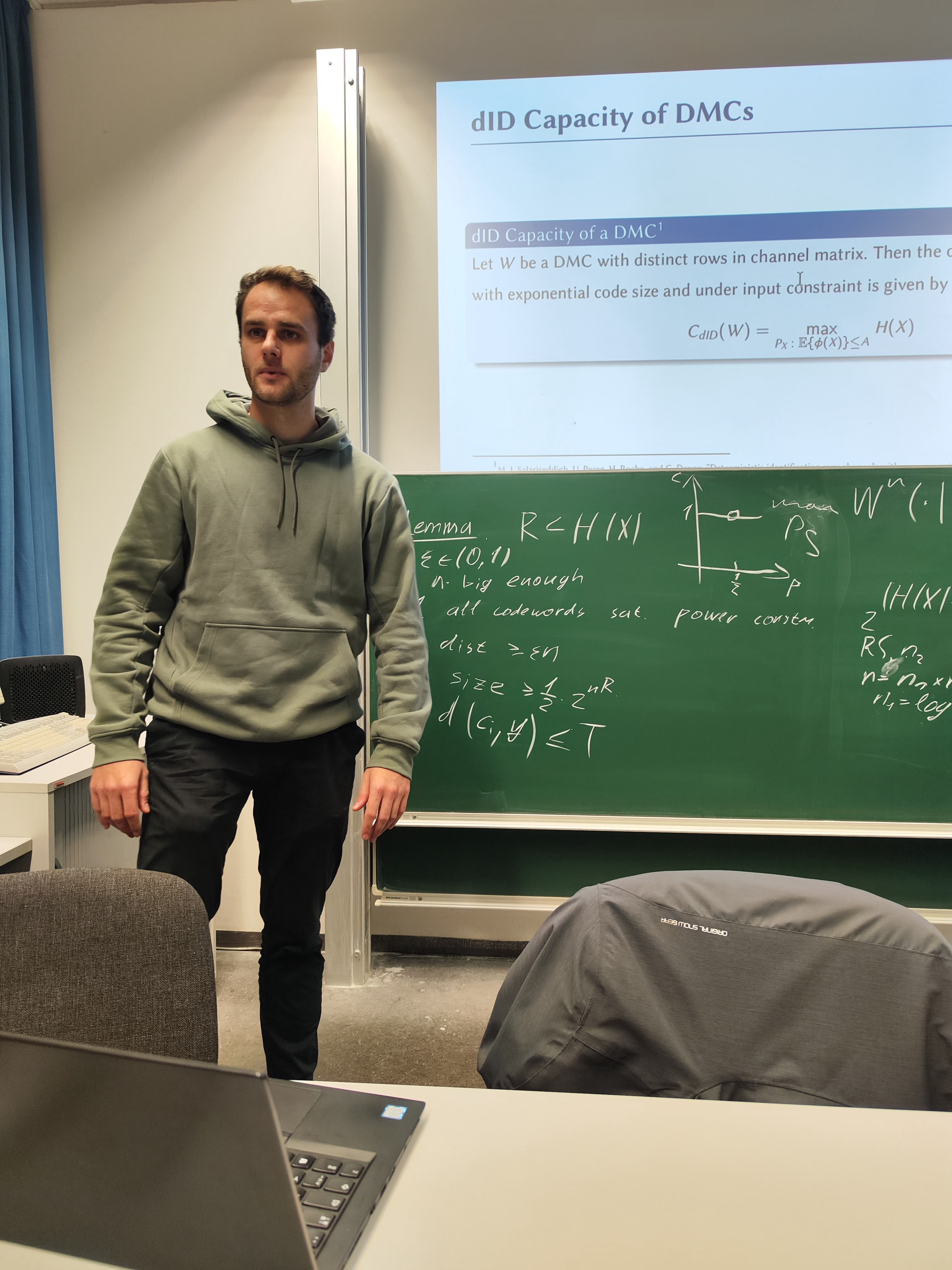}}
\caption{Jonathan Huffmann}
\end{floatingfigure}
Jonathan Huffmann followed on directly from Ilya, building upon the foundation laid by his predecessor. His work centered on the information-theoretic analysis of deterministic identification codes. He presented his ideas, which were based on his joint presentation with Christian Deppe at the workshop in Munich 2023 \cite{huffmann}. This analysis is crucial because it provides bounds on the possible rates at which these codes can operate effectively.
 Jonathan’s analysis revealed some intriguing results, particularly the fact that the derived formulas for these bounds are independent of the signal-to-noise ratio (SNR). This finding is significant because, in many other coding and communication scenarios, the SNR plays a critical role in determining performance limits. The independence from SNR highlights a unique aspect of deterministic identification codes and suggests that their theoretical underpinnings differ fundamentally from other types of codes.
Moreover, Jonathan’s work stands out because the analytical approach and results for deterministic identification codes diverge from those used for transmission codes and randomized identification codes. The methodologies and resulting bounds in these areas are typically influenced by different factors, including the SNR, making the findings on deterministic codes particularly noteworthy.

\newpage

\subsection{Johannes Rosenberger}
\begin{floatingfigure}[r]{6cm}
\mbox{\includegraphics[width=5.5cm]{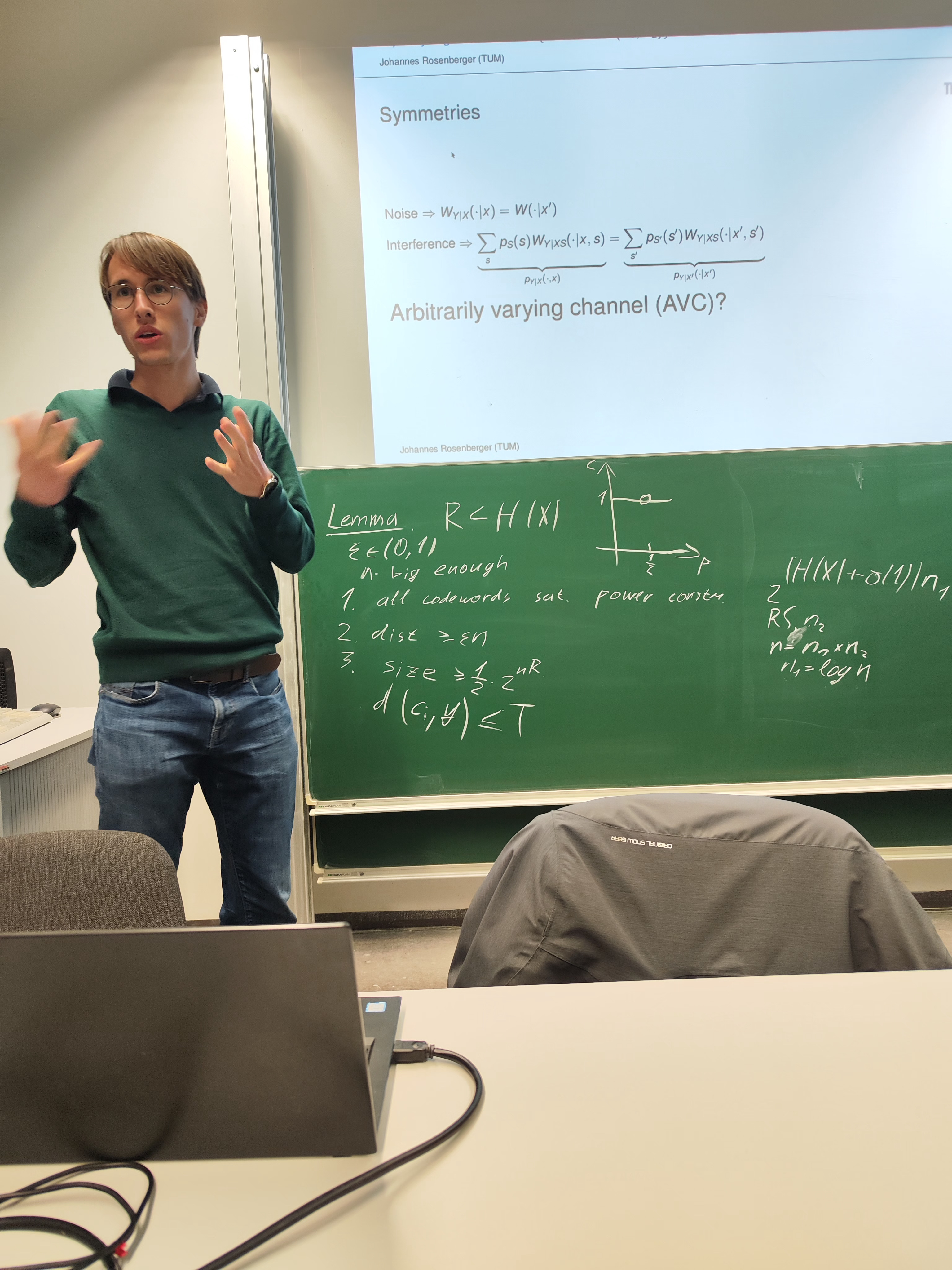}}
\caption{Johannes Rosenberger}
\end{floatingfigure}
Johannes Rosenberger then presented a broader perspective on the topic, delving into the philosophical aspects of post-Shannon theory. He explored the broader implications and the underlying principles of this theoretical framework, using it as a foundation to motivate his research on identification and locally homomorphic channels \cite{rosenberger2024function}.
In his work, Johannes developed the concept of a \textit{locally homomorphic channel}. This type of channel is significant because it enables local operations that preserve certain homomorphic properties, which are essential for various computational tasks. He demonstrated an approximate equivalence between these locally homomorphic channels and codes used for computing functions. This equivalence is a crucial insight, as it bridges the gap between abstract theoretical constructs and practical coding strategies.
Johannes's research led to novel results in the area of $K$-identification \cite{rosenbergerKID}. $K$-identification is a method used in information theory to identify specific messages within a set, and Johannes's findings revealed substantial improvements in the efficiency of this process. These improvements were particularly surprising when compared to more straightforward, naive code constructions, suggesting that his approach could significantly enhance the performance of identification systems.
To illustrate the practical implications of his theoretical results, Johannes provided examples of identification using deterministic encoders \cite{rosenberger2023deterministic}. Deterministic encoders are a specific type of encoding mechanism that maps input data to output codes in a predictable and consistent manner. By applying his new theories to these examples, he was able to showcase the practical advantages and increased rates of identification achievable through his methods.

\newpage

\subsection{Pau Colomer}
\begin{floatingfigure}[r]{6cm}
\mbox{\includegraphics[width=5.5cm]{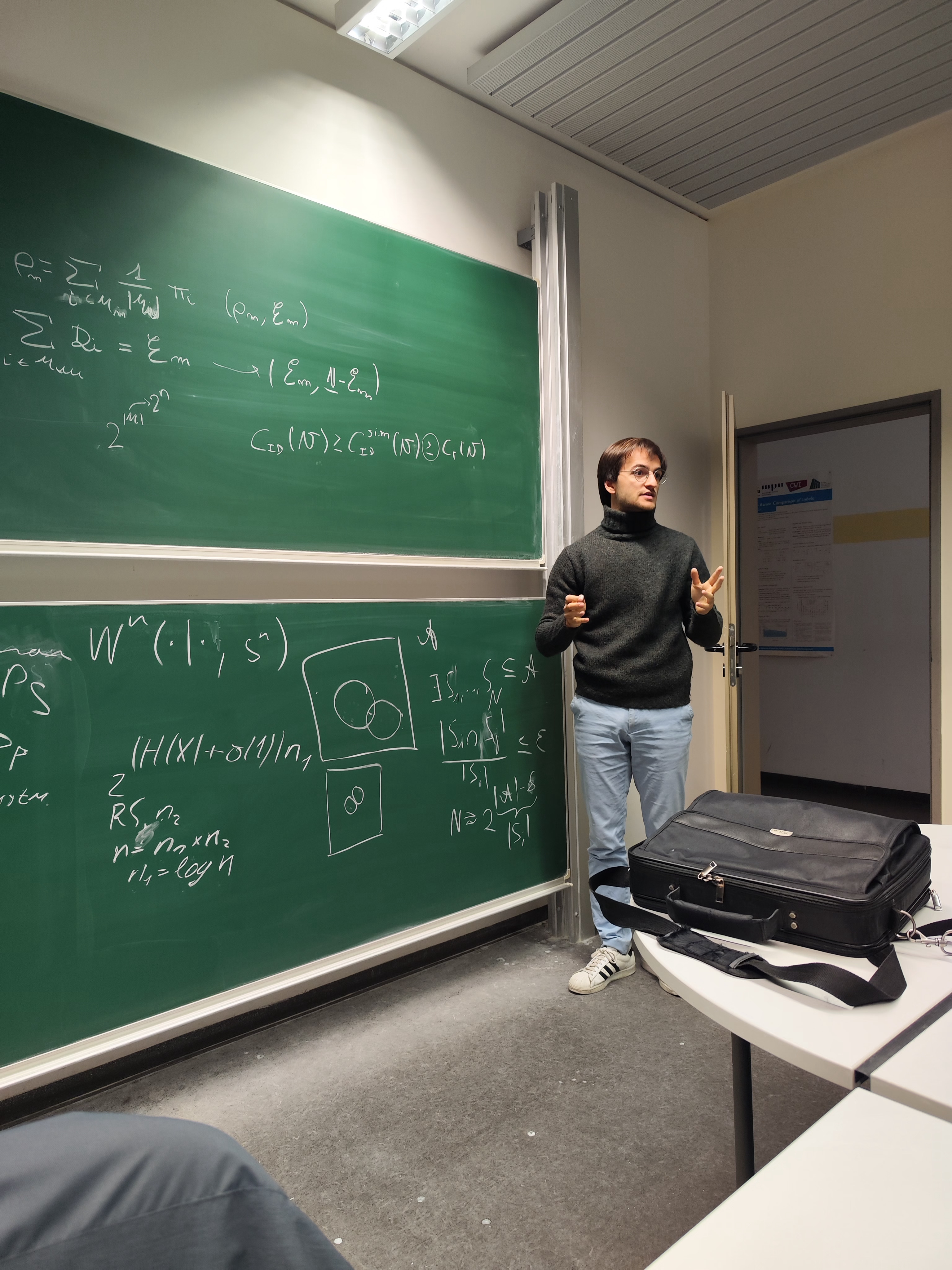}}
\caption{Pau Colomer}
\end{floatingfigure}
The final lecture of the workshop was delivered by Pau Colomer, who faced the formidable challenge of elucidating the intricate topic of quantum message identification \cite{colomer2024zero,colomer2024zeroentropy} and his results in deterministic identification \cite{colomer2024deterministicid,colomer2024deterministic}. The subject quantum communication is particularly complex and less familiar to many attendees, who primarily work in the realm of classical communication and thus lacked a foundational understanding of quantum communication principles.
Pau's task was not only to introduce the basic definitions and concepts but also to bridge the substantial gap between classical and quantum communication paradigms. Despite these hurdles, he adeptly conveyed the core ideas and nuances of quantum message identification. His lecture was thorough and methodical, ensuring that even those without prior knowledge could grasp the essential points.
A significant portion of his presentation focused on the potential extensions and generalizations of message identification within the quantum framework. He explored various theoretical possibilities, delving into how traditional concepts in classical communication could be adapted or redefined in the context of quantum mechanics.
In his current research, Pau is particularly interested in finding an analogue or "twin" for deterministic identification within the quantum domain. This pursuit involves investigating new methodologies and frameworks that can accurately and reliably identify messages, akin to the deterministic processes used in classical communication but leveraging the unique properties of quantum systems.
Throughout his talk, Pau maintained the audience's engagement, despite it being the final session of the workshop. The participants remained attentive and actively participated in discussions, demonstrating their interest and curiosity about the topic. These discussions not only reflected the audience's willingness to delve into the unfamiliar territory of quantum communication but also underscored the effectiveness of Pau's presentation in making a complex subject accessible and intriguing.

\newpage

\section{Social Event}

\begin{floatingfigure}[r]{6cm}
\mbox{\includegraphics[width=5.5cm]{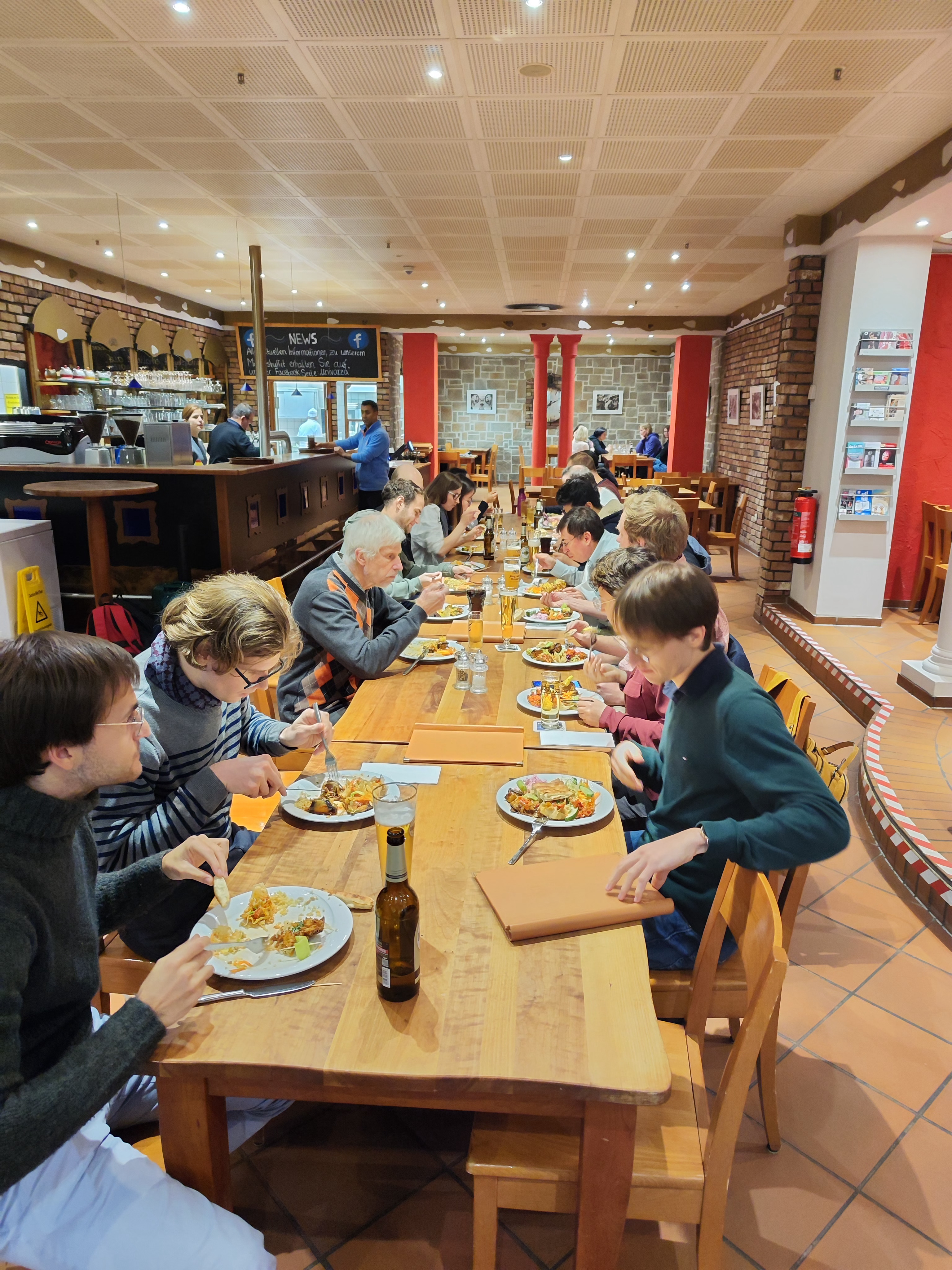}}
\caption{Eating together at a Greek restaurant after the workshop}
\end{floatingfigure}
At the conclusion of the workshop, attendees gathered for a delightful buffet at a Greek restaurant located within Bielefeld University. This setting provided a relaxed and convivial atmosphere, perfect for unwinding after an intensive day of lectures and discussions.
The buffet featured a variety of traditional Greek dishes, including moussaka, souvlaki, spanakopita, and an array of fresh salads and dips such as tzatziki and hummus. The delicious food and warm ambiance encouraged informal conversation and networking among the participants.
Amidst the enjoyment of the culinary offerings, the spirit of scientific exchange remained vibrant. Participants continued to engage in lively discussions about various research topics, with one area of particular focus being the collaborative efforts within the 6G-life project, a cutting-edge initiative between Dresden and Munich.
Researchers from both institutions delved into the specifics of their work on 6G technology, which aims to advance beyond the current capabilities of 5G networks.
The workshop's interactive approach, combining structured presentations with open discussions, was well-received. It provided a valuable platform for knowledge exchange and collaborative learning, setting the stage for the rest of the event.

The satellite workshop held prior to the memorial workshop for Ning Cai was a resounding success. This gathering provided a platform for scholars and researchers to present and discuss various advancements in the field. Notably, many of the results showcased at the event were deeply inspired by Ning Cai's pioneering ideas \cite{DBLP:journals/tit/AhlswedeC06} on message identification. His influential work served as a cornerstone for several presentations, underscoring his enduring impact on the field. The workshop not only honored his legacy but also highlighted the ongoing relevance and application of his theories in current research.

\bigskip

\section*{Acknowledgments}
We extend our gratitude to Jens Stoye from Bielefeld University for generously providing the workshop venues.
The authors acknowledge the financial support by the Federal Ministry of Education and Research of Germany in the program of “Souverän. Digital. Vernetzt.”. Joint project 6G-life, project identification number: 16KISK002. Christian Deppe further gratefully acknowledge
the financial support by the BMBF Quantum Programm QD-CamNetz, Grant
16KISQ077, QuaPhySI, Grant 16KIS1598K, QUIET, Grant 16KISQ093, and NEWCOM Grant 16KIS1005. Yaning Zhao and Christian Deppe were supported by the DFG through Grant DE1915/2-1.

\printbibliography

\end{document}